\documentclass[prd, twocolumn,preprintnumbers, showpacs, groupedaddress]{revtex4}
\usepackage{lineno,hyperref}
\usepackage{hyperref}
\usepackage{color}
\usepackage{float}
\usepackage{bm}
\usepackage{amsmath}
\usepackage{soul}
\modulolinenumbers[5]

\begin{document}
\title{Primordial magnetic field and kinetic theory with Berry curvature}
\author{Jitesh R. Bhatt$^{1}$}
 \email[e-mail:]{jeet@prl.res.in}
\author{Manu George$^{1, 2}$}
\email[e-mail:]{manu@prl.res.in}
\affiliation{%
\centerline{$^{1}$ Physical Research Laboratory, Theory Division, Ahmedabad 380 009, India}\\
\centerline{$^{2}$ Department of Physics, Indian Institute of Technology, Gandhinagar, Ahmedabad 382 424, India}
}
\date{\today}
 \bigskip
 
\begin{abstract}

We consider the generation and evolution of magnetic field in a primordial plasma at
temperature $T\leq 1~MeV$ in presence of asymmetric neutrino background i.e. the number densities of right-handed and left-handed neutrinos are not same. Semi-classical equations of motion of a charged fermion are derived using the effective low-energy Lagrangian. It is shown that the spin degree of freedom of the charged fermion couples with the neutrino background. Using this kinetic equation we study the collective modes of the plasma. We find that there exist an unstable mode. This instability is closely related with the instability induced by chiral-anomaly in high temperature $T\geq 80~TeV$ plasma where right and left-handed electrons are out of equilibrium. We find that at the temperatures below the neutrino decoupling
the instability can produce magnetic field of 10 Gauss in the Universe.
We discuss cosmological implications of the results.

\end{abstract}
\maketitle
\section{Introduction}

There has been a considerable interest in studying magnetic field generation in plasmas
with a parity violating interaction [\cite{SS_04}\cite{Joyce_97}\cite{ Giovannini_98} \cite{Boyarsky_12} \cite{Boyarsky_12a} \cite{Tashiro_12}]. 
Such studies can be potentially useful in understanding the primordial origin of the observed inter galactic 
magnetic fields which is still an open issue (for recent reviews see \cite{Widrow:2011hs}\cite{Yamazaki:2012ab},
\cite{Kanda_15}). They can also be useful in understanding the magnetic field generation in a core-collapsing supernovae or a magnetar [
\cite{Ohnishi_14},\cite{Dornikov_15}]. 
The earlier parity violating mechanism relied on the assumption that there was an excess of right handed electrons over positrons by some  process\cite{Joyce_97}at temperature $T>80~TeV$.  In this case one can define the chiral-chemical potentials for right and left handed particles and they are not equal.For such a situation the right-handed current is not conserved due to the Abelian anomaly.This state can be shown to unstable and the instability can generate a hypercharge magnetic field of $10^{22}~G$. However, at temperature $T<80~TeV$ one can expect  right and left handed electrons in equilibrium and the asymmetry is washed out. But it was shown in Ref.\cite{Boyarsky_12a} that in presence of strong magnetic field the left-right asymmetry can be present till temperature $T\sim 10~MeV$. Similar mechanism for the magnetic field generation in a core-collapsing supernova was proposed 
[\cite{Ohnishi_14} \cite{Dornikov_15}].
In presence of parity violating interaction, the most general expression for photon polariozation tensor $\pi^{ij}$ takes the form [e.g. \cite{Joyce_97}\cite{Boyarsky_12a}\cite{Aka_13}],
\begin{equation}
\pi^{ij}=\pi_L P^{ij}_L+\pi_TP^{ij}_T+\pi_A P^{ij}_A.
\label{ptensor}
\end{equation}

$P$ are the projection operators defined as: $P^{ij}_L=k^ik^{j}/k^2$, $P^{ij}_T=\delta^{ij}-k^ik^j/k^2$ and $P^{ij}_A=i\epsilon^{ijk}k^k/k$ where, $k^i$ is $i$-th component of the propagation vector . 
The first two terms are parity even and very well studied. The coefficient of the axial projection operator will be zero in a parity even theory. Thus the structure and strength of the third term 
depends on the parity odd interactions of the theory. In such  cases the transverse branch of the dispersion relation can split into 
two components \cite{Nieves_89}. 
One of the branches of the transverse dispersion relation become unstable due to
the chiral-imbalance. The instability can be shown to generate magnetic field
in the plasma [\cite{Joyce_97},\cite{Aka_13},\cite{Bhatt_15}].

%
\noindent
Interestingly a similar scenario can exist in the neutrino sector where neutrino
and charged lepton interaction is considered \cite{DS_14}. In this work using the techniques of {\it finite temperature field theory}(FTFT) it was shown that the  magnetohydrodynamics (MHD) is modified in the presence of neutrino asymmetry. The neutrino asymmetry in the plasma was shown to drive the instability which in turn can generate a magnetic field.
Here we would like to note that charged leptons in the plasma were considered in Ref.\cite{DS_14} to be chirally unpolarized. In the low energy regime $m>T$ where, $m$
is the mass of charged leptons the chiral polarization can be washed out as the chirality
flip rated due to collisions $\Gamma\sim {m^2}/{T^2}$. There are some other works where
in a different context neutrino and charged particle interaction was shown to contribute significantly for the anti-symmetric part of the photon polorization tensor
$\Pi_A$ [\cite{Nieves_89}, \cite{Mohanty_98}, \cite{NS_05} see also \cite{SS_04}] over which we shall comment later.


 Recently, there has been an important development in incorporating parity violating
effect within a kinetic theory framework [\cite{SY_12}-\cite{CPWW_13}].
In this approach the Vlasov equation was modified by including the Berry-curvature
term to account for the chirality of charged particles. This modification makes
the kinetic description consistent with the equation of the Abelian anomaly. Moreover, the
 parity odd correlation function calculated using the modified Vlasov
equation is identical with the results of the underlying quantum field theory in the next-to-leading order hard dense loop approximation \cite{SY_13}. However, the neutrino-
charged-fermion interaction was not considered in this work. 
The advantage that the  kinetic theory description offers is that it is much simpler to handle than FTFT and one can have access to a variety of the plasma physics techniques.
Keeping this in mind in the present work we derive the Vlasov equation in the presence
of neutrino asymmetry. It ought to be noted
here that there exist kinetic description of neutrino in a dense medium based upon
Bogoliubov-Born-Green-Kirkwood-Yvon(BBGKY) heirarchy [\cite{VVE_13}, see also \cite{SR_93}] or quantum kinetic description \cite{VFC_14}. Also  the modulation of the neutrino flux by plasma waves
has been considered in Ref.\cite{BBDSS_96}. 
In the present work we  modify the Vlasov equation to
incorporate low energy ($m\geq T$) interaction between the charged leptons and neutrinos.
 The neutrino flux 
is considered to be constant like that in Refs.[\cite{Mohanty_98}-\cite{DS_14}]. 
Using the effective Lagrangian for $\nu e$-interaction \cite{Giunti_07}, we derive a set of  equations
describing  motion of a charged particle in electromagnetic field and the asymmetric neutrino background. We show that the neutrino background couples with spin of the
charged particle.
From these equations the modified Vlasov equation can be obtained. 
It is interesting to note here
that modified Vlasov equations with spin dynamics are also considered in the regular
plasma literature also in a different context [see for example \cite{Brodin_08}]. The magnetics of
a spinning plasma can be extremely richer than the regular plasma \cite{MA_11}. 
We show that the asymmetry in the neutrino background can alter the spin dynamics of the charged particles
and which can be responsible for the generation of magnetic field. We discuss the cosmological consequences of our result.

\section{The equations of motion and the kinetic theory}
Lagrangian density for lepton field interacting with background neutrino is given by,
\begin{equation}
 \mathcal{L}=\bar\psi[i\gamma^\mu\partial_\mu\psi-\gamma_\mu(f^\mu_LP_L+f^\mu_RP_R)-m]\psi
 \label{lagrangian}
\end{equation}
where, m is mass of the lepton, $\gamma^\mu=(\gamma^0, \bm{\gamma})$ are the Dirac matrices and $P_{L,R}=\frac{1\mp\gamma^5}{2}$ are the 
chiral projection operator with $\gamma^5=i\gamma^0\gamma^1\gamma^2\gamma^3 $.  $f^\mu_{L,R}=(f^0_{L,R}, \bm{f}_{L,R})$
are the neutrino currents and they are regarded as an external macroscopic quantities.
An explicit form of $f^\mu_{L,R}$ can be calculated from effective Lagrangian \cite{Giunti_07} 
\begin{equation}
\mathcal{L}_{eff}=[-\sqrt{2} G_F\sum_\alpha\bar\nu_\alpha\gamma^\mu\frac{(1-\gamma^5)}{2}\nu_\alpha][\bar\psi\gamma_\mu(a_L^\alpha P_L+a_R^\alpha
P_R)\Psi]
\label{Leff}
\end{equation}
where, label $\alpha$ denotes neutrino species $\alpha=e,\,\mu,\,\tau$ and  $G_F=1.17\times10^{-11}MeV^{-2}$ is the Fermi constant. The coefficients $a_L^\alpha$ \& $a_R^\alpha$
are given by
\begin{equation}
 a_L^\alpha=\delta_{\alpha,e}+sin^2{\theta_W}-1/2 , a_R^\alpha=
 sin^2{\theta_W}
 \label{aLR}
\end{equation},
with $\theta_W$ being the Weinberg angle. Next, we assume that  $\nu \bar\nu$ form an isotropic background gas.
 This in turn means that in averaging over the
neutrino ensemble, only nonzero quantity will be\, $<\bar\nu_\alpha\gamma^0(1-\gamma^5)\nu>=2(n_{\nu_\alpha}-n_{\bar\nu_\alpha})$. Number densities 
of neutrinos and atineutrinos can be calculated using corresponding  Fermi-Dirac distribution function
\begin{equation}
 n_{\nu_\alpha , \bar\nu_\alpha}=\int \frac{d^3p}{(2\pi)^3}\frac{1}{e^{\beta_{\nu_\alpha}(|\bm p|\mp\mu_{\nu_\alpha} )}+1}
 \label{FDdistribution}
\end{equation}
where $\beta$ is the inverse temperature. Using Eq.\ref{lagrangian}-\ref{FDdistribution} one obtains 
\begin{align}
  f^0_L=&2\sqrt 2 G_F[\Delta n_{\nu_e}+(sin^2\theta_W-1/2)\sum_\alpha\Delta n_{\nu_\alpha})] ,\\  f^0_R=&2\sqrt 2 G_F sin^2
 \theta_W\sum_\alpha\Delta n_{\nu_\alpha}.
\end{align}
Thus the equation of motion obtained from Eq.\ref{lagrangian} can be written as 
\begin{equation}
 i\frac{\partial\psi}{\partial t}=[\bm\alpha\cdot\bm{\hat p}\psi+\beta m-(f^0_LP_L+f^0_RP_R)]\psi.
 \label{eompsi}
\end{equation}
Writing $\psi=\left(\begin {array}{c} \phi \\ \chi \end {array}\right ) $ in the Eq.\ref{eompsi} and following the standard procedure, Hamiltonian
for the  large component of the spinor \cite{Greiner_reqm} can be obtained as,

\begin{equation}
\mathcal{ H}=\frac{1}{2m}\bm{(\sigma\cdot p)(\sigma\cdot p)}+\frac{\Delta f^0}{2m}\bm{(\sigma\cdot p)}+\frac{f^0}{2}+O(f_{L,R}^2)
\nonumber
\end{equation}
where $f^0=f^0_L+f^0_R$ and $\Delta f^0=f^0_L-f^0_R$. In the above equation, we have neglected terms propotional to $G_F^2$.
In the presence of external electromagnetic field,  momentum $\bm p$ has to be 
replaced by $\bm{p-eA}$. Thus the Hamiltonian for charged fermion in interacting with an external electromagnetic
field and background neutrino is given by,

\begin{equation}
  \mathcal {H}=
 \frac{(\bm{p-eA})^2}{2m}-\bm{\mu\cdot B}+eA^0+\frac{\Delta f^0}{2m}\bm\sigma\cdot(\bm p-e\bm A)+\frac{f^0}{2}
\label{H}
\end{equation}
\noindent
where, $\bm\mu=\frac{eg}{4m}\bm\sigma$ is the electron magnetic moment ans $g$ is the Land\`{e} g-factor. The first three terms on the right hand
side  are well known and very well studied in the literature. The fourth and fifth terms are due to the neutrino background. The last term 
might contribute
to  the energy of the system, but it will not enter into the equations of motion as the neutrino background considered to be constant. 
If the neutrino background vary with space and time, this term would modify force 
equation as $\bm F\propto \bm\nabla f=\bm\nabla\psi_\nu^*\psi_\nu$. This force is called pondaromotive force. Such a scenario was studied 
in Ref.\cite{BBDSS_96}, however in their formalism
the fourth term was not considered. 

 In order to find the equation of motion
for a charged particle in an electromagnetic
field and the neutrino background, one can
use Eq.(\ref{H}) and the 
Heisenberg equation $\dot{\hat O}= i[\hat {\mathcal{H}},\hat O]$ and write: 
\begin{equation}
 \bm v=\dfrac{\bm p-e\bm A}{m}+\frac{\Delta f^0}{2m}\bm\sigma
 \label{dotx}
\end{equation}
where we wrote $\dot{\bm x}=\bm v$.
\begin{equation}
 \dot{\bm p}=\frac{e}{m}(\bm p-e\bm A)_k\bm\nabla\bm A_k+\frac{eg}{m}\bm\nabla\bm{(s\cdot B)}-e\bm{\nabla}A^0
 \label{dotp}
\end{equation}
where we have defined $\bm s=\bm\sigma /2$ and
\begin{equation}
 \dot{\bm s}=\mu_B(\bm s\times\bm B)-\Delta f^0(\bm s\times\bm v)
 \label{spin_dyn}
\end{equation}
From the equations \ref{dotx}-\ref{spin_dyn} we get 
\begin{equation}
 \ddot{\bm x}=\frac{e}{m}[\bm E+\bm v\times\bm B]+\frac{e\Delta f^0}{2m^2}(\bm s\times \bm B)+\frac{eg}{2m^2}\bm\nabla(\bm s\cdot\bm B)
 \label{ddotx}
\end{equation}
Next, if $n$ is the particle distribution function in the extended phase-space, then the particle
conservation implies that the phase space density is conserved along the single particle trajectory
i.e. $\partial_t n+\bm{\dot x}\cdot\nabla_{\bm x}n+\bm{\ddot x}\cdot\nabla_{\bm{v}}n+\bm{\dot s}\cdot\nabla_{\bm{s}}n=0$. 
\noindent
Thus with Eqs.\ref{spin_dyn} and \ref{ddotx},
\begin{align}
 \nonumber &\partial_t n+\bm{v}\cdot\bm\nabla_{\bm x}n+[
 \frac{e}{m}(\bm E+\bm v\times\bm B)+\frac{\mu_B}{m}\bm\nabla(\bm s\cdot\bm B)\\  &+
 \frac{\mu_B\Delta f^0}{m}(\bm s\times \bm B)]\cdot\bm\nabla_{\bm v}n+ [\mu
 _B(\bm s\times\bm B)-\Delta f^0(\bm s\times\bm v)]
 \cdot\bm\nabla_{\bm s}n=0
\label{vlasov}
\end{align}
Here we note that when $\Delta f^0=0$, Eq.\ref{vlasov} matches with the spin modified kinetic equation of Ref.\cite{Brodin_08}. 
In order to have a self-consistent set of equations one needs to combine
the Eq.(\ref{vlasov}) with Maxwell's equations with the current density is given by
\begin{equation}
{\bm j}=\sum_i\left[ q_i\int {\bm v}n_id{\bm v}d{\bm s}+3\mu_{Bi}
\nabla\times \int{\bm s}n_id{\bm v}d{\bm s}\right]
\label{mcurrent}
\end{equation}
\noindent
Here, the sum is over particle species $i$ with charge $q_i$. The integral is carried out over three velocity and the two spin variables.
Second term on the right hand side is representing the magnetization current.

 We need to write the Eq.(\ref{vlasov}- \ref{mcurrent}) in curved space-time in order to apply it
to the early Universe scenario. 
In the expanding Universe with Friedmann-Robertson-Walker metric with scale factor $a$, 
the these equations can retain its flat-space  form \cite{Dattmann_93} if we go to the conformal time coordinates defined by  $\eta=\int a^{-1}(t)dt$ with replacements:
$\bm{E}\rightarrow a^{2}E$,$\bm B\rightarrow a^{2}\bm B$, $\bm j\rightarrow a^{3}\bm j$.
In this case Maxwell's equations can be written as:\\
\begin{align}
\nabla\cdot\bm E=&\rho\\
	\partial_\eta\bm B=&-\nabla\times\bm E\\
\nabla\times \bm B=&\bm  j+\partial_\eta \bm E
\label{M2}
\end{align}
where, $\rho$ is the total charge density.

\section{Linear analysis and the dispersion relations}
 Next, we study the linear response analysis of Eqs.(\ref{vlasov}-\ref{mcurrent}). 
For this one writes the charged particle distribution function as 
 $n=n_0({\bm v},{\bm s})+\delta n({\bm x},{\bm v},{\bm s})$, where, 
$n_0=$ and $\delta n$ respectively denote the initial and perturbed parts of the distribution function.
Further, we assume that there is no background electric or magnetic field. We choose $A^0=0$ as a gauge in the subsequent analysis. 
In this situation induced current in the Fourier space  can be
written as $\bm j^i_{\bm k, \omega}=-\pi^{ij}_{\bm k, \omega}A^j_{\bm k, \omega}$,
where, $A^{j}_{\bm k, \omega}$ is the space part of the four
potential $A^\mu_{\bm k, \omega}$. From \ref{vlasov} we get,

\begin{align}
\nonumber
\delta n_{\omega,\bm k}=&\{\frac{e\omega}{T}\bm v\cdot \bm{A_{\omega,\bm k}}- \frac{i\mu_B}{T}\bm{k\cdot v}(\bm{s\times k})\cdot\bm{A_{\omega,\bm k}}
 -\frac{\mu_B\Delta f^0}{T}[(\bm s\cdot \bm{A_{\omega,\bm k}})\bm k \\- &(\bm s\cdot \bm k)\bm{A_{\omega,\bm k}}]\cdot \bm v\}
 \frac{n^0}{\omega-\bm k\cdot \bm v}
 \label{delta_n}
\end{align}
Here  we have assumed that the space-time dependence of the perturbations to be $e^{-i(\omega t-\bm k\cdot \bm x)}$.
Substituting  equation \ref{delta_n} into  \ref{mcurrent}, we get  
\begin{align}
	\nonumber \pi^{ij}_{\omega,\bm k}=& e\int\frac{ d\bm v~ d\bm{s}}{4\pi}~\delta(s-1)\{\frac{e\omega}{T}v^iv^j-\frac{i\mu_B}{T}(\bm k\cdot\bm v) 
	v^i(\bm s\times\bm k)^j
	\\\nonumber
- &\frac{\mu_B\Delta f^0}{T}[(\bm k\cdot\bm v)v^is^j-(\bm s\cdot\bm k)v^iv^j]\}
	\frac{n^0}{\omega-\bm k\cdot \bm v}\\ \nonumber+ &
	3i\mu_B\int\frac{ d\bm v~ d\bm{s}}{4\pi}~\delta(s-1)(\bm k\times \bm s)^i\{\frac{e\omega}{T}v^j-\frac{i\mu_B}{T}(\bm k\cdot\bm v)(\bm s\times\bm k)^j
	\\ \nonumber-&\frac{\mu_B\Delta f^0}{T}[(\bm k\cdot\bm v)s^j-(\bm s\cdot\bm k)v^j]\}
	\frac{n^0}{\omega-\bm k\cdot \bm v}.
\end{align}
Functions $\pi_L=P_L^{ij}\pi^{ij}_{\omega,\bm k}$, 
$\pi_T=P_T^{ij}\pi^{ij}_{\omega,\bm k}$ \& 
$p_A=P_A^{ij}\pi^{ij}_{\omega,\bm k}$ in Eq.\ref{ptensor}, now
can be calculated using the above equations together 
with the definition of the projection operators and
they are given below:
\\
\small{
\begin{align}
\pi_L=&~~\frac{e^2\omega}{Tk^2}\int\frac{ d\bm v~ d\bm{s}}{4\pi}~\delta(s-1)\frac{(\bm k\cdot\bm v)^2n^0(v,\bm s)}
{\omega-\bm k\cdot \bm v}\\
 \nonumber
\pi_T=&~~e\int \frac{ d\bm v~ d\bm{s}}{4\pi}~\delta(s-1)\{\frac{e\omega}{2T}[v^2-\frac{(\bm k\cdot \bm v)^2}{k^2}]-\frac{i\mu_B}{T}\bm {k.v(s\times k).v}\\
\nonumber &-\frac{\mu_B\Delta f^0}{T}[\bm{(k.v)(s.v)}-\bm{(k.s)}v^2]\}
\frac{n^0(v,\bm s)}{\omega-\bm k\cdot \bm v}\\
&+3i\mu_B\int\frac{ d\bm v~ d\bm{s}}{4\pi}~\delta(s-1)\{\frac{e\omega}{T}(\bm k\times\bm s).\bm v+\frac{i\mu_B}{T}\bm k.\bm v
(\bm s\times\bm k).(\bm s\times \bm k)\\ \nonumber &+\frac{\mu_B\Delta f^0}{T}\bm s.\bm k (\bm k\times\bm s).\bm v\}\frac{n^0(v,\bm s)}
{\omega-\bm k\cdot \bm v} 
\end{align}
}
\small{
\begin{align}
	\pi_A=& -\frac{e\mu_B}{2kT}\int\frac{d\bm v~d\bm s}{4\pi}\delta(s-1)\epsilon_{ijl}k_lv_i(\bm s\times\bm k)_j
\frac{\bm k.\bm v n^0(v,\bm s)}{\omega-\bm k\cdot \bm v}\\
\nonumber
 &+\frac{ie\mu_B\Delta f^0}{2kT}\int\frac{d\bm v~d\bm s}{4\pi}\delta(s-1)\epsilon_{ijl}k_lv_i[\bm k.\bm v s_j-\bm s.\bm k v_j]
\frac{n^0(v,\bm s)} {\omega-\bm k\cdot \bm v}\\ \nonumber
&+\frac{3e\mu_B\omega}{2kT}\int\frac{d\bm v~d\bm s}{4\pi}\delta(s-1)\epsilon_{ijl}(\bm k \times \bm s)_i v_j k_l
\frac{n^0(v,\bm s)} {\omega-\bm k\cdot \bm v}\\ \nonumber
&-\frac{3\mu_B^2\Delta f^0}{2kT}\int\frac{d\bm v~d\bm s}{4\pi}\delta(s-1)\epsilon_{ijl}k_l(\bm k\times \bm s)_i[\bm k.\bm v s_j-\bm s.\bm k v_j]
\\ \nonumber &\times\frac{n^0(v, \bm s)} {\omega-\bm k\cdot \bm v}.
\end{align}
}
Equation for the transverse branch of the dispersion relation:
\begin{equation}
	\omega^2-k^2=\pi_T\pm\pi_A.
\label{dispersion_g}
\end{equation}
After integrating over the spin degrees of freedom, equations for $\pi_L$, $\pi_T$
and $\pi_A$ are written as follows:
\begin{align}
\pi_L(\bm k,\omega)=&\frac{e^2\omega}{k^2T}\int d\bm v(\bm k\cdot \bm v)^2\frac{n_0(v)}{\omega-\bm k\cdot \bm v},\\ 
\pi_T(\bm k,\omega)=&\int d\bm v\left[\frac{e^2\omega}{2T} \left(v^2
-\frac{(\bm k\cdot\bm v)^2}{k^2}\right)-\frac{\mu_B^2k^2}{T}
(\bm k\cdot \bm v) \right]\frac{n_0(v)}{\omega-\bm k\cdot \bm v},\\
\pi_A(\bm k,\omega)=& \frac{\mu_B^2\Delta f^0k}{T}\int d\bm v~~
(\bm k\cdot \bm v) \frac{n_0(v)}{\omega-\bm k\cdot \bm v},
\end{align}
\noindent
where, $\omega_p^2=n_0e^2/m$ is square of the plasma frequency.
We emphasize here that in Refs.[\cite{Mohanty_98}, \cite{NS_05}] photon propagation in a neutrino gas in presence of finite chemical potential was considered. But in these work the role of
unstable modes was not analyzed. In Ref.\cite{Mohanty_98} how the dispersion Eq.(\ref{dispersion_g}) can influence the rotation of the plane of polarization of electromagnetic waves over the cosmological distances was considered. However, in this work no real charged-leptons were considered. In Refs
[\cite{NS_05},\cite{DS_14}] it was shown that
the neutrino-charge lepton interaction can contribute
very significantly to $\pi_A$.  Further, we note that the instability occurs in the quasi-stationary regime\cite{Aka_13} i.e. $|\omega|\ll k$.
Thus by integrating over velocity and keeping only linear terms in $\omega/k$, we get
\begin{align}
\Pi_L(\bm k,\omega)=&O\left(\omega^2/k^2\right)\\
\pi_T(\bm k,\omega)=&\omega_p^2\left[\left(\frac{g^2k^2}{4mT}\right)
-i\sqrt{\frac{2\pi m}{T}}(\frac{\omega}{k})\right]\\
\pi_A(\bm k,\omega)=&\omega_p^2\left(\frac{g^2}{4mT}\right)\Delta f^0 k\left[1-i\sqrt{\frac{2\pi m}{T}}(\frac{\omega}{k})\right]
\label{pi_a}
\end{align}

In the quasi-stationary limit the dispersion relation for the transverse mode
can be written as
 \begin{equation}
 	\omega_\pm=-i\sqrt{\frac{T}{2\pi m}}\left[{\frac{k^3}{\omega_p^2}
\pm \frac{g^2}{4mT}\Delta f^0k^2}
\right].
\label{dispersion}
 \end{equation}
 
%
Here clearly the root $\omega_{-}$ gives instability
if the condition $ k< \omega_p \frac{\omega_p}{T}\frac{\Delta f^0}{m}$ is satisfied.
$k$ dependence of Eq.(\ref{dispersion}) is similar to the instability found
in Ref.\cite{Aka_13} for the chiral plasma of massless particle. It should
be noted here that in the present work we work we have not considered the chirally polarized
charged fermions like in Refs.[\cite{Joyce_97}, \cite{Aka_13}].

From Eq.(\ref{dispersion}) one can easily find value of wave vector $k_{max}$ for which the instability is maximum. Thus we write:
\[k_{max}\sim\left(\frac{4\omega_p^2}{3mT}\right)\Delta f^0.\]
\noindent
In presence of collision we have to add 
collision frequency $\nu_c$ term  in Eq. \ref{vlasov}. In the collision dominated regime i.e. $\nu_c\gg \omega,k$ Eq.\ref{pi_a} can be written as:
\begin{equation}
	\pi_A\approx -\frac{2\omega^2_p}{mT}\Delta f^0 k. 
\end{equation}
\noindent
Here we note that our expression of $\pi_A$ is has  similar form
like the one reported in Ref.\cite{DS_14}. In this regime one can write the
dispersion relation for the transverse mode
\begin{equation}
	\omega_\pm=-i\frac{k^2}{\sigma}\pm\frac{2i}{\sigma}\frac{\omega^2_p}{mT}\Delta f^0 k
\label{hydrodisp}
\end{equation}
\noindent
where, $\sigma=\frac{\omega^2_p}{\nu_c}$.
 whose $+$ branch is unstable if $k< \frac{2\omega^2_p\Delta f^0}{mT}$ is
satisfied and the maximum growth rate occurs at $k_{max}=\frac{\omega^2_p}{mT}\Delta f^0$.
One can notice here that the condition for instability and value of $k_{max}$
are similar in both collision-dominated and the collision less cases.
$k$ dependence of Eq.(\ref{hydrodisp}) is similar to one in a parity-violating
magnetohydrodynamical limit as in Ref.[\cite{Joyce_97}, \cite{Bhatt_15}].
The instabilities that we have found in Eqs.(\ref{dispersion},\ref{hydrodisp})are qualitatively similar to $\alpha$-effect in the Solar physics where a large scale magnetic field get self-excited due to violation of mirror symmetry
in the turbulence. Interestingly in Ref.\cite{SS_04} this effect was considered in the context of neutrino plasma
interaction. In this work the asymmetric neutrino background
was considered to have inhomogeneity scale smaller than the 
magnetic field. In our work the neutrino background is homogeneous. Further the dispersion in the collision dominated regime i.e. Eq.(\ref{hydrodisp}) is similar to the modes that one might obtain from eqn.(9) in Ref.\cite{SS_04}. Importantly in our formalism we have $\alpha
-$effect in the collisionless regime which was not found in Ref.\cite{SS_04}. We can estimate the $\alpha -$ effect from magnetic conductivity equation:

\begin{equation*}
	\partial_t \bm B\sim \eta\nabla^2\bm B+\alpha\nabla\times\bm B
\end{equation*}

For the collisionless case, using the Eqns. (27-29) we write $j^i=-(\Pi_L P^{ij}_L+\Pi_T P^{ij}_T+\Pi_A P^{ij}_A)A^j$ and obtain expression for $i\bm k\times\bm B_{\bm k}=\omega_p^2[\frac{k^2}{mT}\bm B_{\bm k}+\sqrt{2\pi m/T}\frac{i\omega\bm B_{\bm k}}{k}]+i\frac{\Pi_A}{k}\bm k\times\bm B_{\bm k}$. Next, Using Maxwells equations $\nabla\times\bm B=\bm J$ and $\nabla\times\bm E=-\partial_t\bm B$ one can write the magnetic conductivity equation for this case:

\begin{equation*}
	-i\omega\bm B_{\bm k}=\frac{(\omega_p^2/mT-1)k^3\bm B_{\bm k}+i\Pi_A\bm k\times\bm B_{\bm k}}{\omega_p^2\sqrt{2\pi T/m}}
\end{equation*}
 
This gives $\alpha=\sqrt{\frac{2\pi}{mT^3}}\Delta f^0 k$, this is a new result . Similarely for the collision dominated case we obtain, $\alpha=\frac{2\omega_p^2}{\sigma}\frac{\Delta f^0}{mT}$ which is similar to that found in ref.\cite{DS_14}.

Next, we consider the only the electron-neutrino species and thus 
Following Ref.\cite{KT_book} number density of electrons $n_0$ for the case
$T<m$, can be written as $n_0\sim 2\left(\frac{mT}{2\pi}\right)^{(3/2)}$.
The neutrino asymmetry in the Universe is constrained by the BBN abundance
of $^4$He. The neutrino number asymmetry depends only on the electron-neutrino
degeneracy parameter $\xi_{\nu e}$ \cite{Serpico_05}. For our purpose we use
the neutrino asymmetry  $\Delta n_{\nu e}\approx 0.061 \xi_{\nu e}T^3_\gamma$
\cite{Geng_07}, where $T_\gamma$ is the photon temperature and it is related
to current CMBR temperature $T^\prime_\gamma$ by the formula 
$T_\gamma=(1+z)T^\prime_\gamma$ and $z$ is the red shift. Now $\Delta f^0$
can be estimated to be $\Delta f^0=\sqrt 2G_F\Delta n_{\nu e}$. By considering
the plasma just after the neutrino decoupling at temperature less than 1 MeV
and $z\sim 10^9$ one finds $k^{-1}_{max}\sim 10^4 $cms with $\xi_{\nu e}=0.072$.
At this instant the horizon size can be estimated to be $H^{-1}\sim 10^{14}$cms \cite{KT_book}.
One can also
estimate strength of the produced magnetic field as follows: When the instability saturates, there are no macroscopic motion.
From the spin dynamics equation, magnitude
we camn write 
$ B \sim \frac{\Delta f^0}{
\mu_B} v_{th}$, where $B=|\bm B|$ and also  we have replaced $\bm v$ by the thermal velocity. Using the above
estimate $\Delta n_\nu$ one can find
$B\sim 1.9\times 10^{-14}$ (MeV)$^2\sim  10$ Gauss. Similar values of
$B$ can be estimated if $B\sim k_{max}A$ and $A\sim T$. 
%

 In conclusion we have developed kinetic theory for the spin plasma in
the neutrino background. 
In this formalism the charged particles are not
in their chirally-polarized states. 
The normal modes of the spin modified equation in the neutrino background
are shown to give the dispersion
relation similar to the chiral-plasma where the charged particles are considered to be massless.
It was shown that the asymmetry between densities of right-handed 
and left-handed neutrinos of the
background can induce an axial part in the photon polarization tensor and 
give rise to unstable modes related with the 
so called $\alpha -$effect.
Further, we have shown that the $\alpha -$ effect can exist for both 
the collisionless and the collision-dominated regimes and it can lead to generation of magnetic field.
We have also shown that 
the length scale associated with the unstable modes ($k_{\max}^{-1}\sim 10^4 $cms) at the time of the neutrino decoupling in the early Universe.
We have also estimated the strength of magnetic field generated through above discussed mechanism.
\section*{Acknowledgements}
We sincerely thank Profs. A. Joshipura, S. Rindhani and S. Mohanty for their enlightening discussions and invaluable comments.

\end{document}